\documentclass{article}

\usepackage{microtype}
\usepackage{graphicx}
\usepackage{subfigure}
\usepackage{listings}
\usepackage{booktabs} % for professional tables
\usepackage{hyperref}

\usepackage[accepted]{sysml2019}

\sysmltitlerunning{Auto-Vectorizing TensorFlow Graphs}

\begin{document}

\twocolumn[
\sysmltitle{Auto-Vectorizing TensorFlow Graphs: Jacobians, Auto-Batching and Beyond}

% unless you've provided the [accepted] option to the sysml2019


\sysmlsetsymbol{equal}{*}

\begin{sysmlauthorlist}
\sysmlauthor{Ashish Agarwal}{google}
\sysmlauthor{Igor Ganichev}{google}
\end{sysmlauthorlist}

\sysmlaffiliation{google}{Google Inc, Mountain View, California, USA}

\sysmlcorrespondingauthor{Ashish Agarwal}{agarwal@google.com}
\sysmlcorrespondingauthor{Igor Ganichev}{iga@google.com}

\sysmlkeywords{Machine Learning, SysML, TensorFlow, Auto-batching, Jacobians, Vectorization}

\vskip 0.3in

\begin{abstract}

We propose a static loop vectorization optimization on top of high level dataflow IR used by
frameworks like TensorFlow. A new statically vectorized parallel-for abstraction is provided
on top of TensorFlow, and used for applications ranging from auto-batching and per-example gradients,
to jacobian computation, optimized map functions and input pipeline optimization. We report huge
speedups compared to both loop based implementations, as well as run-time batching adopted by the
DyNet framework.
\end{abstract}
]


% The \sysmlEqualContribution command is standard text for equal contribution.

\printAffiliationsAndNotice{}  % leave blank if no need to mention equal contribution
% \printAffiliationsAndNotice{\sysmlEqualContribution} % otherwise use the standard text.

\section{Introduction}
\label{Introduction}
Today's machine learning applications operate on huge multi-dimensional arrays, or \emph{tensors}, and typically run 
algorithms involving large chunks of parallelizable computations. The inherent vectorization opportunities
are typically leveraged by writing highly optimized libraries of common operations, or \emph{kernels}, for each platform. 
Common frameworks like TensorFlow \cite{DBLP:journals/corr/AbadiABBCCCDDDG16} and PyTorch \cite{paszke2017automatic} 
contain hundreds of such kernels, which can be viewed as the
instruction set for writing ML programs. Using this instruction set of kernels,
and optionally other control flow constructs,
most frameworks define a high level \emph{IR} (intermediate representation)
which is either constructed explicitly, or implicitly via tracing JIT (e.g. \cite{frostig18jax}).
This IR either gets interpreted, or JIT or AOT compiled after going through different layers of optimizations. 
\cite{xla2017,wei2017dlvm,DBLP:journals/corr/abs-1802-04799}.

We propose a static loop vectorization optimization that operates on this high level IR.
In contrast to CPU instruction sets, these kernels often allow fusing arbitrary number
of scalar instructions, where the count may not even be known statically. Thus instead of statically unrolling a fixed number
of loop iterations, one can conceptually unroll the full loop and fuse the contained instructions, often completely getting rid of the loop, especially for embarrassingly parallel computation. 
This process can then be repeated in cases with nested control flow constructs.

We have implemented a library that performs vectorization on TensorFlow 
Graph IR. The optimization triggers during the graph construction phase
and can currently vectorize more than a hundred different types of kernels.
The user provides the body of a parallel-for loop and gets back symbolic
tensors representing the result of a vectorized version of that loop.

This construct of vectorization-optimized parallel-for loop lends itself to 
many interesting applications. A commonly desired use case is auto-batching
\cite{DBLP:journals/corr/LooksHHN17,NIPS2017_6986} wherein the user writes model code on
a single example and then applies that function to each input in a batch. Such code is more
intuitive and simpler,
especially for models involving dynamic computation graphs.
Auto-batching can be implemented by putting the forward pass
of the model inside a parallel-for loop with batch size number of iterations.
A variant of this is getting per-example gradients, which can be computed by putting
both the inference and the gradient computation inside a parallel-for loop. 
Increasingly, models involve more deeply nested sequential and parallel loops, like
iteration over sequence inputs \cite{Gers:2000:LFC:1121912.1121915},
Markov Decision Processes, sampling or particle filters for stochastic models, adaptive computation \cite{DBLP:journals/corr/Graves16}, and many more, providing a huge
opportunity for applying these techniques.

Another commonly desired application is computing \emph{jacobians}.
When the output of a network is a vector or tensor, jacobian involves computing gradients of
each scalar value in the output. This is a powerful mathematical object but unfortunately not supported natively
by TensorFlow which implements jacobian-vector products instead.
Iterative approaches often end up being slow. Putting gradient computation of each output element in a parallel-for loop and
vectorizing it enables more efficient computation. Additionally it enables
many cases that don't currently work in TensorFlow, like jacobians of dynamic RNNs
or \emph{hessians} of non-scalar outputs. Our jacobian implementation has enabled new research work like \cite{2018arXiv180602215P} and \cite{golub}.

\section{Related Work}
\label{related}

Auto vectorization is a well known technique applied by compilers like GCC and LLVM and has 
been well researched in literature.
\cite{Nuzman:2006:AID:1133255.1133997} proposed compiler transformations that support vectorization in the presence of interleaved data.
A polyhedral model \cite{Trifunovic:2009:PGL:1636712.1637770} can be used to estimate the performance 
impact of the different loop transformations and vectorization strategies. 
\cite{Barik:2010:ESV:1934902.1935022} use dynamic programming for selecting vector instructions.
Techniques for performing whole function vectorization is discussed  in \cite{Karrenberg:2011:WV:2190025.2190061}.
ispc \cite{ispc} proposed an SPMD programming model for graphics workload.
Halide \cite{Ragan-Kelley:2013:HLC:2491956.2462176} provides a model for separation of algorithm from 
schedule, allowing  stochastic search over space of schedules. 

Most of these approaches
operate at much lower level of abstraction and hence miss out on domain specific optimizations,
like leveraging linear algebra identities. Frameworks like TensorFlow XLA \cite{xla2017},
TVM \cite{DBLP:journals/corr/abs-1802-04799}
and DLVM \cite{wei2017dlvm} follow a staged compilation approach, and can
fuse operations  at high level. Our work operates at a similar or higher abstraction level
and proposes applying loop vectorization along with adding a new parallel-for user abstraction.

Auto-batching is a special case of vectorization and has been used by TensorFlow Fold
\cite{DBLP:journals/corr/LooksHHN17} and DyNet \cite{NIPS2017_6986}. Both these approaches
use run-time or \emph{dynamic batching}, which is very flexible and can work even when the kernels are dispatched
dynamically from a host language like Python. However dynamic batching can potentially incur large overheads,
both in trying to identify and fuse operations, as well as in collating the inputs and slicing the outputs 
when necessary. Additionally control flow dependent on computation can cause the dispatch to stall and hence
fusion cannot work across it.

A static auto vectorization approach can reduce the dispatch overheads,
vectorize across nested control flow and allow additional post-vectorization optimizations.
In dynamic environments, the technique could be applied to JIT compiled traces \cite{frostig18jax}.
We recently became aware of static auto-batching effort \cite{matchbox} for PyTorch 
which works by rewriting the Python AST. We believe that static analysis
on Python could potentially miss out on optimization opportunities given that
types may not be known statically, or it would have to delay it to run-time
which may incur additional overheads. By performing static auto vectorization
on strongly typed high level IR, we are able to achieve large speedups across
a range of applications.

\newcommand{\gammahat}{\hat{\gamma}}
\newcommand{\xhat}{\hat{X}}
\newcommand{\pfor}{\texttt{pfor}}

\section{Compilation Details}
\label{compile}

\subsection{Programming Model}
\label{model}

We assume a dataflow programming model, extended to handle state and control flow, similar to the one
used for TensorFlow \cite{Yu:2018:DCF:3190508.3190551}. A directed cyclic graph represents a program or computation.
Its nodes represent \textit{operations}, or primitive functions. Its edges represent data flowing between operations.
Edges incident into a node represent its inputs, and edges incident
out of the node represent its outputs. Both inputs and outputs are ordered. The data on edges can be either immutable
\textit{tensors} or mutable \textit{variables}. Both tensors and variables are multi-dimensional arrays of primitive types.

Nodes get ready to run when all their inputs are ready. When multiple nodes are ready, they can be executed
in any order or in parallel.
Besides the edges representing data dependencies,
the model includes edges representing \textit{control dependencies}. A control edge
from node $\eta_1$ to node $\eta_2$ enforces that node $\eta_2$ is executed after node $\eta_1$.

The graph can have cycles due to loops and may have special
operation nodes to implement different control flow semantics. In addition, we assume that all operation types as well as the types of all input and output tensors are known at compile time. Full or partial shapes may be known statically as well.

Even though we focus on a dataflow based programming model, for the purpose of exposition, we will write
imperative pseudocode and examples. These snippets should be seen as building the dataflow graph. For example
$Z = matmul(X, Y)$ represents a subgraph with a single node representing the operation $matmul$ with two inputs edges, $X$ and $Y$, and a single output edge $Z$. Similarly, imperative constructs
like \texttt{for}, \texttt{while}, \texttt{if-then-else} should be seen as constructing the corresponding control
flow constructs in the graph. In addition, we will use \texttt{parfor} to represent a \textit{parallel-for} loop. 
Its semantics will be discussed in more detail in \S\ref{problem}.

\subsection{Notation}
\label{notation}
We use the following notation. $\gamma$ represents program segments, corresponding to subgraphs
in the overall dataflow graph. Optionally, we use notation $\gamma({args})$ to 
identify the set ${args}$ as some of the inputs of interest to $\gamma$.
$\gammahat$ represents the compiled version of
$\gamma$, where the goal and semantics of compilation will be described in subsequent sections.
$\eta$ will represent a node in $\gamma$ and $\hat{\eta}$ is its compiled version in $\gammahat$. It
may be a single node or a  sub-graph. $op_{\eta}$ is the operation for the node $\eta$.
For any $X$ which is a tensor (or a sequence of tensors), $\hat{X}$ represents
the \emph{vectorized} version of $X$. A vectorized version means that $\hat{X}$ 
stores $n$ different versions of $X$ in some layout. 
$X_j$, where $j$ is an integer, or a vector of integers, refers to a row (or rows) $j$ along the first
dimension and either \textit{gathers} them into a new tensor, or references them as a mutable variable. If $X$ is a sequence, the definition applies recursively for each element.
However if $\hat{X}$ represents a vectorized version of $X$, then $\hat{X}_j$
represents the $j^{th}$ version of $X$, if $j$ is scalar. If $j$ is a vector, $\hat{X}_j$ represents a vectorized entity that \textit{subsets} the underlying versions to be $j$, i.e. only includes versions $j$.

\subsection{Problem Statement}
\label{problem}

Given a dataflow graph $\Gamma$ represented by the parallel-for loop of the form

\begin{algorithmic}
  \PARFOR{$i=0$ {\bfseries to} $n-1$  }
    \STATE $\gamma(i)$   
  \ENDFOR
\end{algorithmic}

where $n$ is a scalar integer tensor whose value may only be known at run-time, and $\gamma(i)$ is a set of instructions that may depend
on the loop variable $i$, as well as any global state, the goal of the compilation process is to generate a
new dataflow graph, $\gammahat(n)$, that is functionally equivalent to $\Gamma$, and that tries to get
rid of or reduce the scope of the \texttt{parfor} loop. 

A \texttt{parfor} loop is defined similar to a \texttt{for} loop with the difference that 
the iterations don't need to 
run in sequential order. Instead the output and side-effects of the execution should be the 
same as under a \textit{SIMD} execution model.

A SIMD execution of
 $\Gamma$  involves dispatching the nodes of $\gamma(i)$ in lock step for
all \texttt{active} iterations, which are the iterations of the loop that are
executing that operation. This set starts off as being $\{0,...,n-1\}$. Whenever
a control flow block construct is encountered, the set is updated to be the set
of iterations that continue execution of that block and other inactive iterations
stall, waiting for the block to finish. For \texttt{parfor}, 
note that we don't require that the instructions actually be run in 
lock-step across all active iterations, but that the global state when the execution is done be the same as after a SIMD execution.

\subsection{Compiling Stateless Operations}
\label{stateless}

We start with the case where the result of \texttt{parfor} is the same
under any ordering or interleaving of the execution of the different iterations.
(i.e. no data dependencies between any of the iterations, and no side-effects).
\S\ref{stateful} will describe adding support for stateful operations.

Here we provide examples on how to compile simple stateless graphs,
building towards our conversion algorithm in \S\ref{greedy}. 
We will directly list $\gamma(i)$ in our examples below and assume 
that it is wrapped in a \texttt{parfor} as illustrated in \S\ref{problem}.
$\gammahat$ will be the vectorized 
version and the last expression will be assumed to be the returned value. 
Shapes of the inputs are listed on the first line.

An important aspect demonstrated here is how we leverage
\texttt{loop invariance} of tensors beyond what traditional approaches
to loop invariant code motion achieve. This will involve generating 
different code based on what combinations of input values are 
loop invariant.

For the purpose of examples here and later,
we adopt the following layout for vectorized tensors. We  assume here
that shapes are fixed across iterations.
Let $X$ have shape $[s_0,...s_r]$. Then we can layout $\xhat_j$ by stacking all the 
versions of $X$ sequentially. Given that,
shape of $\xhat$ is $[n, s_0,...s_r]$ and $\xhat_j = \xhat[j], 0 \le j < n$, corresponds to the value of
$X$ in iteration $j$. However if $X$ is loop invariant, $\xhat$ has same shape
as $X$ and represents the value of $X$ in all iterations, i.e. we avoid
unnecessary stacking.
For cases where shape of $X$ is different across iterations,
more complicated book-keeping is needed, and
we ignore such cases for the examples here.
Also note that when shape information is statically available it can be used for 
more efficient code generation. Else the compiler generates code conditioned
on run time shapes.

The common operation of gathering the $i^{th}$ row of a tensor with first dimension $n$ can be vectorized as follows.
\begin{algorithmic}
\INPUT{$X: [n, x]$}
\STATE $\gamma(i): X[i]$
\STATE $\gammahat(n): X$
\end{algorithmic}

In the more general case, gathering rows can be vectorized as a slicing operation
on $X$.

\begin{algorithmic}
\INPUT{$X: [m, x]$}
\STATE $\gamma(i): X[i]$
\STATE $\gammahat(n): X[0:n, ...]$
\end{algorithmic}

Component-wise operations can invoke the original operation.

\begin{algorithmic}
\INPUT{$X: [n, x, y], Y: [n, x, y]$}
\STATE $\gamma(i): X[i]  + Y[i]$
\STATE $\gammahat(n): X + Y$
\end{algorithmic}

Typically such operations support broadcasting \cite{broadcasting}
that semantically tiles the inputs along 
corresponding dimensions to get the shapes to match before 
component-wise operations are performed. Shapes are left extended with $1$ to get ranks to 
match, and corresponding dimensions should have equal values or at least one of them should be $1$. 

Compilation may need extra reshapes to makes sure broadcasting works for generated
code. Also note below how it handles loop invariant input $X$ and gets away
without having to tile it $n$ times. Here $reshape$ is assumed to reshape the 
input without copying the underlying data.

\begin{algorithmic}
\INPUT{$X: [y, z], Y: [n, z]$}
\STATE $\gamma(i): X  + Y[i]$
\STATE $\gammahat(n):$
  \STATE $X1 = reshape(X, [1, y, z])$
  \STATE $Y1 = reshape(Y, [n, 1, z])$
  \STATE X1 + Y1
\end{algorithmic}

Now assume $matmul$ performs matrix multiplication and $batch\_matmul$
takes two lists of matrices (as 3-D tensors) and performs matrix multiplication
on the corresponding values in the lists.

\begin{algorithmic}
\INPUT{$X: [n, x, y], Y: [n, y, z]$}
\STATE $\gamma(i):  matmul(X[i], Y[i])$
\STATE $\gammahat(n): batch\_matmul(X, Y)$
\end{algorithmic}

The above is effectively the same as the uncompiled version. However if one of the 
inputs is loop invariant, this can be optimized based on the mathematical properties
of matrix multiplication.
\begin{algorithmic}
\INPUT{$X: [n, x, y], Y: [y, z]$}
\STATE $\gamma(i):  matmul(X[i], Y)$
\STATE $\gammahat(n):$
  \STATE $X1 = reshape(X, [n * x, y])$
  \STATE $R = matmul(X1, Y)$
  \STATE $reshape(R, [n, x, z])$
\end{algorithmic}

Next assume $conv2D$ performs convolution on the input using the passed in filter.
We will assume that the operation performs padding of the input to ensure that the
output shape for each channel is the same as input.
\begin{algorithmic}
\INPUT{$X: [n, b, h, w, c1], F: [k1, k2, c1, c2]$}
\STATE $\gamma(i):  conv2D(X[i], F)$
\STATE $\gammahat(n):$
  \STATE $X1 = reshape(X, [n * b, h, w, c1])$
  \STATE $R = conv2D(X1, F)$
  \STATE $reshape(R, [n, b, h, w, c2])$
\end{algorithmic}

Next we look at reductions. Here $reduce\_sum$ reduces the input by performing
sum reduction along the passed in axes. The generated code renumbers the axes
and calls the same reduction.
\begin{algorithmic}
\INPUT{$X: [n, x, y, z]$}
\STATE $\gamma(i):  reduce\_sum(X[i], [1, -1])$
\STATE $\gammahat(n):reduce\_sum(X, [2, -1]$
\end{algorithmic}

Renumbering of axes works for a lot of other operations. Here
is an example of $concat$ which concatenates a list of tensors along
a particular axis.
\begin{algorithmic}
\INPUT{$X: [n, x, z], Y: [n, y, z]$}
\STATE $\gamma(i):  concat((X[i], Y[i]), 1)$
\STATE $\gammahat(n): concat((X, Y), 2)$
\end{algorithmic}

In our implementation described in \S\ref{implementation}, we added support for
converting more than 100 different operations.
As illustrated by above examples, lot of these conversions involve
calling the original operation $op_\eta$ with additional reshape and occasional transposes. 
In other cases, $op_\eta$ is flexible
in the length of one dimension, often a \textit{batch} and sometimes a \textit{channel} dimension,
and operates independently on each 
slice along that dimension. In such cases, one could fold the first dimension of $\xhat$ into the 
that dimension, run $op_\eta$, and transpose and reshape the outputs back as needed.
Renumbering axes was another common approach. 

Note that whether a particular input is loop invariant or not is independent of the operation
being converted and is a property of how the dataflow graph was structured. Hence when implementing
a converter, one may need to consider all possible  combinations of loop invariance of its inputs.
However based on typical graphs compiled, and based on which of those
combinations can be expressed efficiently using the set of operations available to the framework,
one can leave most of these combinations unoptimized by falling back to the loop based implementation 
as written above. This fallback provides a path for incrementally baking in more optimizations over time.

\subsection{Greedy Algorithm}
\label{greedy}

We provide a greedy algorithm for the problem stated in \S\ref{problem}.
Given the dataflow graph corresponding to 
$\gamma(i)$, we first convert it to a directed acyclic graph by converting each control flow block
into a single node. Next we traverse this new dataflow graph in topological order, and for each node $\eta$, 
generate a new set of nodes, $\hat{\eta}$, that efficiently implement the functionality of running $\eta$ in a
\texttt{parfor}. Any control dependencies of $\eta$ are mirrored for each node in
$\hat{\eta}$. Each output tensor $X$ of $\eta$ is mapped to a new tensor (or sequence of tensors) $\xhat$
in $\hat{\eta}$ that stores all the versions of $X$ across all iterations
using some chosen layout. 

Code generation is done by using a registry of \emph{converters} keyed by the signature of the 
operation, $op_\eta$ corresponding to the node $\eta$. See \S\ref{stateless} and \S\ref{stateful}
for discussion  on these converters. If the node represents a  control flow 
block in the original graph, special logic is employed
which traverses the nodes corresponding to this block, strips out control flow operations and extracts
sub graphs that correspond to semantics blocks of that control flow, like the \texttt{body} and 
\texttt{condition} of loops,
or the \texttt{condition}, \texttt{then} and \texttt{else} blocks for conditionals, etc. These blocks are 
recursively converted and new control flow code is generated as detailed in \S\ref{conditional} and
\S\ref{loops}.

\subsection{Compiling Conditionals}
\label{conditional}

Next we look at compiling code with conditionals. Consider the code segment below where $T$ and $R$
are tensors or sets of tensors, and $\gamma^{pre}$, $\gamma^{cond}$, $\gamma^{then}$ and
$\gamma^{else}$ are code segments representing dataflow subgraphs.

\begin{algorithmic}
  \PARFOR{$i\gets0$ {\bfseries to} $n-1$  }
    \STATE $T, C \gets \gamma^{pre}(i)$ 
    \IF {$C$}
      \STATE $R \gets \gamma^{then}(i, T)$
    \ELSE
      \STATE $R \gets \gamma^{else}(i, T)$ 
    \ENDIF
  \ENDFOR
\end{algorithmic}

Below is a functionally equivalent code segment that gets rid of the \texttt{parfor}. The idea
is to evaluate the \texttt{if} condition for all the iterations, then compute the indices of the 
iterations, $I^{then}$  and $I^{else}$, that would go into the \texttt{then} and \texttt{else} blocks
respectively. Given these indices, we compile $\gamma^{then}$ and $\gamma^{else}$ to work for only 
those particular iterations. The compilation process needs to care about some details. 
Firstly, it needs to subset any vectorized tensors to only those active indices. Secondly any references
to $i$ in $\gamma^{then}$ should vectorize to $I^{then}$ (similarly $I^{else}$ for $\gamma^{else}$). Also,
the number of iterations should be set to be $n^{then}$ and $n^{else}$ respectively. Given these
things are taken care of, the converter for conditionals can call the compilation module recursively
to convert $\gamma^{then}$ and $\gamma^{else}$.

\begin{algorithmic}
\STATE $\hat{T}, \hat{C} \gets \gammahat^{pre}(n)$
\STATE $I^{then} \gets \{j: \hat{C}_j$ is True$\}$
\STATE $n^{then} = length(I^{then})$ 
\STATE $I^{else} \gets \{j: \hat{C}_j$ is False$\}$
\STATE $n^{else} = length(I^{else})$ 
\IF {$n^{then} > 0$}
  \STATE $\hat{T}^{then} \gets \hat{T}_{I^{then}}$
  \STATE $R^{then} \gets \gammahat^{then}(n^{then}, \hat{T}^{then})$
\ENDIF
\IF {$n^{else} > 0$}
  \STATE $\hat{T}^{else} \gets \hat{T}_{I^{else}}$
  \STATE $R^{else} \gets \gammahat^{else}(n^{else}, \hat{T}^{else})$
\ENDIF
\STATE $\hat{R} \gets Scatter((I^{then}, I^{else}), (R^{then}, R^{else}))$
\end{algorithmic}

where $Scatter$ stitches back the partial results from the two branches based on the indices
that each branch processed. An interesting case is when loop condition $C$ is loop invariant.
In such cases, the subsetting operations can be skipped for performance reasons.

\subsection{Compiling Nested Loops}
\label{loops}
Now let us consider the case of loops nested inside the \texttt{parfor}. If that loop is itself a \texttt{parfor}
loop, then the compiler can be invoked to convert it first. In case of sequential \texttt{while}
and \texttt{for} loops, corresponding converters need to be invoked. If these loops are deeply nested, each 
converter will in turn invoke compilation recursively which will lead to converting these loops inside out.

Here is a \texttt{while} loop nested inside a \texttt{parfor}. As before
$T$ and $R$ are tensors or sets of tensors, and $\gamma^{pre}$, $\gamma^{cond}$ and $\gamma^{body}$ are code segments representing dataflow subgraphs.

\begin{algorithmic}
  \PARFOR{$i \gets 0$ {\bfseries to} $n-1$  }
    \STATE $R, T \gets \gamma^{pre}(i)$ 
    \WHILE {$\gamma^{cond}(i, R_i, T)$}
      \STATE $R \gets \gamma^{body}(i, R, T)$
    \ENDWHILE
  \ENDFOR
\end{algorithmic}

Below is the functionally equivalent code that gets rid of the \texttt{parfor}. The overall idea
is to generate another \texttt{while loop} where, in each iteration, we keep track of the indices, $I$, of
all \texttt{parfor} loops that are still active, and run the condition and body, $\gammahat^{cond}$
and $\gammahat^{body}$, on only those indices. Similar to the conditional case, this involves subsetting all the  vectorized inputs of those blocks to the set $I$, and having the recursive compilation be aware of the  list and count of active iterations. 
Also, a more optimized implementation can be done for the case
where the output of $\gamma^{cond}$ is loop invariant. 

\begin{algorithmic}
\STATE $\hat{R}, \hat{T} \gets \gammahat^{pre}(n)$
\STATE $done \gets false$
\STATE $I \gets [0,...,n-1]$
\STATE $l \gets n$
\WHILE {not $done$}
  \STATE $C \gets \gammahat^{cond}(l, R_I, \hat{T}_I)$
  \STATE $I' \gets \{j: C_j$ is true$\}$
  \STATE $I \gets I_{I'}$
  \STATE $l \gets length(I)$
  \IF {$l$ is $0$}
    \STATE $done \gets true$
  \ELSE
    \STATE $R_I \gets \gammahat^{body}(l, R_I, \hat{T}_I)$
  \ENDIF
\ENDWHILE
\end{algorithmic}

\subsection{Compiling Stateful Operations}
\label{stateful}

Most of discussion till now has ignored stateful operations. Turns out even handling stateless
operations is sufficient for optimizing many practical applications. Nonetheless we extend
our approach for stateful operations.

For cases where $\hat{\eta}$ touches (reads or writes) any mutable state, 
we adopt the following safety mechanism. Firstly, any accesses to state start
only after all the inputs are ready. Secondly, all state accessed anywhere by
$\hat{\eta}$ is protected (using mutex) for the entire duration of any state accesses.
Lastly, the first output should be produced only after all accesses to state are done. 
This allows us to simulate a similar ordering imposed when executing $\eta$. See
\S\ref{discussion} for how this assumption helps prove equivalence of the converted code. 

Operations that mutate state may often not be SIMD compatible. Even assigning a value to a mutable variable
is not compatible, unless the value assigned is loop invariant. Given that a user is putting this
call in a \texttt{parfor} loop, there are multiple options to handle it. Firstly, it could 
be raised as an error. Secondly, assuming these calls are intended and useful, 
they could be overloaded with special semantics. Thirdly, if instead of SIMD semantics, user
intended sequential semantics, 
the conversion could be implemented for sequential output. Choice of these options could be driven 
by context (e.g. we were optimizing a sequential loop vs parallel loop) as well as special flags
(raise error vs overload semantics for SIMD case).

An example of such a case is random number generation.
The call is not SIMD compatible since each call involves reading and updating some internal 
mutable state. However in SIMD setting it could be desirable to  change the shape passed
to the generator, increasing the rank by 1. Note that the output may or may not be the same as 
under sequential semantics based on the implementation of the generator.

We now discuss some common cases of stateful converters. First we consider operations that are idempotent. Typical examples include get-or-create a 
named mutable tensor or reading the value of a mutable tensor. Given the idempotent nature
these can be run once and the outputs can be marked as loop-invariant. 
Next we consider operations that are commutative and associative. Examples include adding or subtracting
a value into a mutable tensor. Efficient implementation for such operations is done by first
reducing all the updates and then applying the reduced value to the mutable state. 

\subsection{Discussion}
\label{discussion}

Greedy conversion can miss out on many optimization opportunities. For example, data layout
used when converting a node could be better chosen based on the nodes consuming the output. Operations generated for one node could be potentially fused or swapped with subsequently generated ones. Some of these optimizations
can be done independently in subsequent rewrite passes. 

Vectorization can have trade-offs. Memory utilization is typically larger and memory 
constraints may inhibit conversion. Vectorization may not even speed up the compute. Additional
cost models, heuristics and search may be needed to figure out when and how much to vectorize.
However given that we expect a staged compilation, we contend that downstream optimizations 
can use domain specific optimizations,
or tiling and loop reordering to relieve the memory pressure.
In fact, in our experience the generated code is close to what
users can, or already are, writing by hand, and vectorized parallel-for loop becomes just 
another abstraction for expressing high level computation. This further supports the need for subsequent optimizations.

Next we argue that $\gammahat(n)$ is functionally equivalent to SIMD execution of $\Gamma$. 
To establish equivalence, we need to show that given any valid execution ordering of nodes in
$\gammahat(n)$, there is an execution ordering of nodes in $\gamma(i)$ whose SIMD execution produces the same output, and vice versa.
We provide a sketch below. A formal proof, and dealing with stateful nested control flow will 
be left as out of scope of this paper.

First consider an execution ordering of $\gamma(i)$. A valid execution order in $\gammahat(n)$
can be constructed by mapping each node $\eta$ to some valid execution ordering 
of nodes in $\hat{\eta}$. To see why this is a valid execution order of $\gammahat(n)$,
first note that  given our greedy conversion, there is an isomorphism between $\gamma(i)$ and 
$\gammahat(n)$ which maps each node $\eta$ in the former to the sub-graph $\hat{\eta}$ in the latter.
Secondly, functional equivalence of $\eta$ and $\hat{\eta}$ follows from the correctness requirement
of the converter for $op_\eta$. Thus, the given ordering in $\gammahat(n)$ simulates the SIMD execution
of $\gamma(i)$.

Next consider the reverse direction, i.e. we are given an execution ordering in $\gammahat(n)$.
We can similarly map each sub-graph $\hat{\eta}$ to a node $\eta$ in $\gamma(i)$, but
there are some tricky bits here. Firstly execution of $\hat{\eta}$ could be interleaved with execution 
of other nodes in $\gammahat(n)$. This is due to the non-determinism inherent in dataflow execution
where it can run ready nodes in any order and even concurrently. Secondly, there could be multiple operations in $\hat{\eta}$ that touch mutable state at different times and we need to make sure $\eta$ sees the same value for
the state. To get an execution ordering, we first define a value  $T_{\hat{\eta}}$ as follows.
If $\hat{\eta}$ is stateless $T_{\hat{\eta}}$ is the time when the last input 
(including control dependency edges) to $\hat{\eta}$ got ready. For the case $\hat{\eta}$ accesses some state, we 
define $T_{\hat{\eta}}$ as the time the first read/write of some state actually happened inside $\hat{\eta}$.
An equivalent execution ordering of $\gamma(i)$ can now be created by replacing $\hat{\eta}$ with $\eta$ and 
ordering this sequence by $T_{\hat{\eta}}$.

To see why this ordering works, we first show that it is a valid execution order for $\gamma(i)$. If 
$T_{\hat{\eta}_1} < T_{\hat{\eta}_2}$, data or control dependencies from
$\hat{\eta}_2$ to $\hat{\eta}_1$ are ruled out since all outputs of $\hat{\eta}_2$ are produced after
$T_{\hat{\eta}_2}$ while all inputs of $\hat{\eta}_1$ are ready before $T_{\hat{\eta}_1}$ (see
\S\ref{stateful}). Given our 
isomorphism, this similarly means that dependency from $\eta_2$ to $\eta_1$ is ruled out, thus ensuring
validity of execution order. Next we argue why this ordering of nodes $\eta$ simulates the execution of
$\gammahat(n)$. To see that, note that accesses to state inside $\hat{\eta}$
are atomically done and in the order defined by $T_{\hat{\eta}}$. Ordering $\eta$ in the same
order thus allows us to enforce that the state evolves in the same way in $\Gamma$.

\section{Implementation}
\label{implementation}

Using the ideas mentioned in \S\ref{compile}, we have implemented vectorization support in TensorFlow.
We had the option of implementing this as an optimizing rewrite in the TensorFlow C++ runtime.
However we chose to keep this in the
Python frontend, as detailed below. This allowed us to build this external to TensorFlow and independent of the
framework internals. In addition, for simplicity we chose to directly expose a \texttt{parfor} abstraction
to the user instead of transparently optimizing sequential loops. 

We provide a new Python function, \texttt{pfor}, with the following signature: \texttt{pfor(loop\_body\_fn, iters)}.
Here \texttt{loop\_body\_fn} is a Python function that takes a scalar integer tensor,
representing the loop variable, as input, and returns a nested structure of tensors. \texttt{iters}
is a scalar integer tensor representing the number of 
\texttt{pfor} iterations to run.
On being called, \texttt{pfor}  returns a new set of graph nodes whose semantics
is to runs the dataflow graph represented by \texttt{loop\_body\_fn} \texttt{iters} times,
passing the values $0, 1,...iters-1$ to the different iterations, and stacking the outputs returned
by these iterations. Here is an example.
\begin{lstlisting}[language=python,basicstyle=\ttfamily,columns=fullflexible]
a = tf.random_uniform([10, 20])
b = tf.random_uniform([10, 20])

def body(i):
  a_i = tf.gather(a, i)
  b_i = tf.gather(b, i)
  return a_i + b_i, a_i - b_i

# Equivalent to a + b, a - b
output = pfor(body, 10)
\end{lstlisting}
Here \texttt{body} returns two tensors with shape $[20]$ that respectively represent the sum and the difference
of the $i^{th}$ rows of $a$ and $b$. The semantics of this code is to run this function $10$ times, passing
values $0,...,9$ and then stacking the outputs, returning two tensors, each with shape $[10, 20]$. Given
our vectorization process, the call to \texttt{pfor} above
returns tensors $a + b$ and $a - b$.  

Internally, the way it works is that the call to \texttt{pfor}
first makes a single call to the function
\texttt{body} to create a graph with four nodes (two \texttt{Gather}, one \texttt{Addition}, one
\texttt{Subtraction}). Next the compiler is invoked which walks this graph using the greedy procedure
describer earlier, and calls the converters for each node in the graph. Here the converters for Gather
will return \texttt{a} and \texttt{b} respectively, since the vectorization process will essentially
invert gathering each row. Next, converter for \texttt{Addition} will return \texttt{a + b} since that is the 
vectorization of adding the corresponding rows of its inputs. Similarly for \texttt{Subtraction}. Finally
\texttt{pfor} will return these two symbolic tensors.

Most of our work has focused on cases where the shapes of all inputs and outputs are loop invariant,
and we have implemented support for more than 100 TensorFlow kernels, allowing us to vectorize models
like convolutional networks and dynamic LSTMs.
Shape invariance assumption allows us to use the vectorization layout we described in \S\ref{stateless},
which stacks all the underlying values along the first dimension.

We have also built some initial prototypes that handle shape variance, by 
padding the inputs to maximum shape and keeping track of the actual shapes of the underlying components.
In initial experiments with networks involving convolutions, relu
and dense layers on top of inputs with ragged shapes,
we were able to generate code that closely matched the performance of hand written code.

\section{Applications}
\label{applications}

\subsection{Jacobians}
\label{jacobians}

TensorFlow doesn't provide native support for computing jacobians efficiently. Instead
it implements jacobian-vector products.
It is easy to see that jacobian computation is essentially a parallel-for, where each 
iteration computes the gradient of one scalar value in the output tensor. However implementing this using
a \texttt{tf.while\_loop} is slow and doesn't always work. In particular, TensorFlow's \texttt{tf.while\_loop}
gradient computation involves popping tensors out of a TensorFlow stack data structure which is populated during
the forward pass. This means that these gradient computations cannot be repeated multiple times as the stack will
already be empty after the first gradient call is finished. Given this, jacobian computation of \texttt{tf.while\_loop}
(and in turn, \texttt{tf.nn.dynamic\_rnn}) is not supported by TensorFlow.
For the same reason, hessians (jacobian of jacobian) of
non-scalar outputs doesn't work since that will involve jacobian of a \texttt{tf.while\_loop} generated by 
the first jacobian call.

We provide a \texttt{jacobian} function, implemented using \pfor, that is much faster than
an iterative approach (\S\ref{benchmark:jacobians}). Given deep learning experiments can run for hours to days, these speedups alone
make new research feasible. In addition, it enables the cases mentioned above, i.e.,
jacobians of \texttt{tf.while\_loop} and \texttt{tf.nn.dynamic\_rnn}, as well as hessians of non-scalar outputs.

Our support for jacobian has already been used by multiple researchers. For example, \cite{2018arXiv180602215P}
build upon efficient jacobian computation to create a framework for computing eigenfunctions of linear operators
via stochastic optimization, and use it for unsupervised feature generation on video data.
\cite{golub} use our implementation to build a framework for finding and analyzing the fixed points of RNNs.

\subsection{Auto-Batching}
\label{autobatching}

With auto-batching, users write their model code with a batch size of one, and then run that code across different inputs in a batch.  The auto-batching framework makes it run faster by fusing similar operations. See \S\ref{related} for
a comparison of static and dynamic batching approaches and
\S\ref{bench:auto_batching} for benchmarks.

Auto-batching can be implemented by invoking the forward pass (or some part) of the model in a \texttt{pfor} loop. 
Here is some example code.
\begin{lstlisting}[language=python,basicstyle=\ttfamily,columns=fullflexible]
def body(i):
  image, label = input_fn()
  prediction = model_fn(image)
  loss = loss_fn(prediction, label)
  return loss

losses = pfor(body, batch_size)
optimizer.minimize(losses)
\end{lstlisting}
As mentioned earlier in \S\ref{implementation}, most of our current implementation assumes shape invariance, but we have some initial results generating code that automatically pads and unpads ragged tensors. This can enable
auto-batching popular networks, like \emph{Transformers} \cite{2018arXiv180703819D}.
A harder case is efficient auto-batching
of tree and graph traversals and tree RNNs (e.g. \cite{DBLP:journals/corr/GilmerSRVD17,DBLP:journals/corr/TaiSM15}), where padding may not be very efficient. On top
of that, TensorFlow's support for recursive data structures and functions is somewhat limited which may
require resorting to iterative traversals and specialized representations. Getting a good trade-off of
user experience and performance with such models is an open research problem.

\subsection{Per-Example Gradients}
The per-example gradients are regular gradients of the loss with respect to the variables, but the contributions from different examples in a batch are kept separate, not summed. Per-example gradients allow
more sophisticated optimization strategies (e.g. \cite{2015arXiv151106481A}), but TensorFlow users today either need to run at batch size of 1, or perform special surgery on the generated graphs to compute per-example gradients. (\cite{2015arXiv151001799G}).

Computing per-example gradients can be seen as a special case of auto-batching where one puts the gradient computation inside the \texttt{pfor} loop as well, and provides a cleaner and more efficient method for the same. Extending the example from
\S\ref{autobatching}
\begin{lstlisting}[language=python,basicstyle=\ttfamily,columns=fullflexible]
def body(i):
  image, label = input_fn()
  prediction = model_fn(image)
  loss = loss_fn(prediction, label)
  return gradients(loss)
  
per_eg_grads = pfor(body, batch_size)
\end{lstlisting}
\S\ref{benchmark:per_eg} shares some benchmark numbers for computing per-example gradients.

\subsection{Optimizing Map Functions And Input Pipelines}

\texttt{tf.map\_fn} runs a given function, \texttt{fn}, over all row slices of a tensor,
or a set of tensors, and stitches back the generated outputs. This can be directly implemented 
using \pfor. Here is a sample implementation for the simple case of a single tensor input.
\begin{lstlisting}[language=python,basicstyle=\ttfamily,columns=fullflexible]
def pfor_map_fn(f, x):
  return pfor(
    lambda i: f(tf.gather(x, i)),
    tf.shape(x)[0])
\end{lstlisting}
Similarly \texttt{tf.data.Dataset} based pipelines provide a \texttt{map} function that transforms each of the 
elements of the \texttt{Dataset}. This is generally followed later by a call to \texttt{batch}. These
calls can be swapped and the function passed to \texttt{map} can then be vectorized using the given batch size.

We benchmarked some toy input pipelines and simple \texttt{map\_fn} calls. Using \texttt{pfor} 
sped these up one to two orders of magnitude, in-line with results seen in \S\ref{benchmarks}.

\begin{figure*}[t]
%\vskip 0.2in
\begin{center}
\centerline{\includegraphics[width=1.0\linewidth]
    {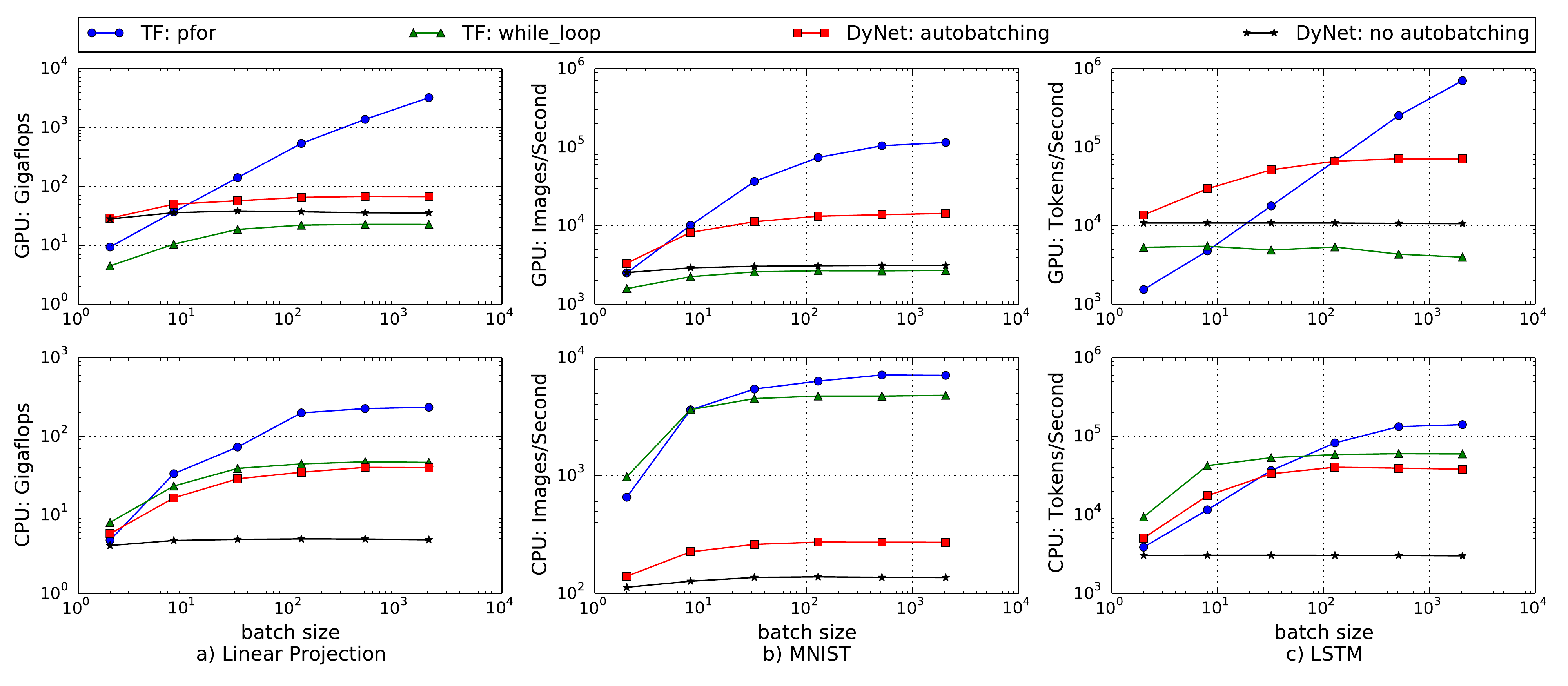}}
\vspace{-10pt}
\caption{Auto-batching 3 models using DyNet and TensorFlow on GPU and CPU. Each column shows performance of a different model. Top row shows performance on GPU and bottom row on CPU. The x-axis is the batch size, i.e. the number of input examples that are auto-batched (in \texttt{pfor} and \texttt{DyNet} with auto-batching) or iterated
over (for \texttt{tf.while\_loop} and \texttt{DyNet} without auto-batching). The y-axis is the achieved throughput.}
\label{fig:autobatch}
\end{center}
\vskip -0.2in
\end{figure*}

\section{Benchmarks}
\label{benchmarks}

\subsection{Setup}
\label{setup}
Experiments were run on a 6 core Intel Xeon E5-1650 3.60GHz CPU with 64GB of RAM and a NVIDIA Maxwell Titan X GPU. We used two models in multiple experiments below. MNIST's architecture is described in \cite{mnist}. It is a stack of two conv-relu-maxpool blocks followed by a linear-relu-dropout-linear block. Inputs are batches of $28x28$ images and output has shape [10]. The LSTM model we used is a single-layer unidirectional RNN based on the LSTM cell described in \cite{lstm}.
Inputs are sequences of 128 dimensional vectors.
LSTM state size is 256 except if mentioned otherwise (e.g. in \S\ref{benchmark:jacobians}).

\subsection{Auto Batching}
\label{bench:auto_batching}

We compare the inference performance of 3 models implemented in different 4 way:
using \texttt{pfor} auto-batching, using \texttt{tf.while\_loop} to loop over inputs in a batch,
and using DyNet with and without auto-batching enabled. Experiments are run with and without GPU support and
reported in Figure \ref{fig:autobatch}. Throughput (measured as gigaflops or images/tokens per second)
is reported as function of batch size.

\subsubsection{Linear Projection}
\label{linear}

We first study applying linear projection on input data. Inputs are randomly generated 768-dimensional vectors of floats.
Projection matrix is a constant $768x768$ matrix of floats.
Figure \ref{fig:autobatch}a reports throughput measured in Gflops against batch size.

We notice that DyNet performs better than TensorFlow at small batch sizes, especially on GPU. In this regime,
the computation time is dominated by fixed overheads, which appear lower for DyNet.
For moderate to large batch sizes, \texttt{pfor}-based
implementation outperforms the other three, by 1 to 1.5 orders of magnitude. The DyNet auto-batching based
implementation does not scale as well at higher batch sizes likely because the auto-batching is performed at
runtime and its cost is proportional to the batch size.

\texttt{tf.while\_loop} does well on CPU since it supports running multiple loop iterations in parallel,
hence utilizing multiple CPU cores. In our implementation, DyNet without auto-batching is driven from
a single Python thread and fails to utilize multiple CPU cores. It is likely possible to use multiple Python
threads to achieve higher performance without auto-batching.

\subsubsection{MNIST}
\label{bench:mnist}
Figure \ref{fig:autobatch}b reports the number of images processed
by the MNIST model per second as we vary the batch size. Overall trends look similar to those of the linear projection benchmark in \S\ref{linear}.
One noticeable difference is that on CPU \texttt{tf.while\_loop}'s performance scales better
than in the linear projection benchmark. This is likely because the MNIST model requires more
computation per iteration and execution overheads become less significant compared to it. \texttt{pfor}
still outperforms the other three for moderate and large batch sizes.

\subsubsection{LSTM}

Figure \ref{fig:autobatch}c reports tokens per second processed by an LSTM  as we vary
batch size. Input sequence lengths for LSTM are sampled uniformly at random between 1 and 100, inclusively. The LSTM is implemented by iterating to the actual length for each sequence. We neither perform padding of the inputs, nor do we iterate to the maximum sequence length in a batch.

Here \texttt{pfor} needs relatively larger batch sizes
to outperform other implementations and the margins are narrower.
This is likely caused by the following two factors. Firstly, our
implementation of vectorizing \texttt{tf.while\_loop} still has considerable overheads,
which we are working on optimizing. Secondly, given that sequence lengths are randomly chosen, iterations of the 
sequential loop generated by \texttt{pfor} progressively operate on smaller batches. 
Smaller batch sizes have lower hardware utilization and also smaller speedup from vectorization
compared to other strategies. The overall
speedup is an expectation over the speedups at different batch sizes and hence lower than
the speedup for linear and MNIST models, which have a fixed batch size throughout an experiment.

\subsection{Per-Example Gradients}
\label{benchmark:per_eg}

\begin{figure}[t]
%\vskip 0.2in
\begin{center}
\centerline{\includegraphics[width=1.0\linewidth]
    {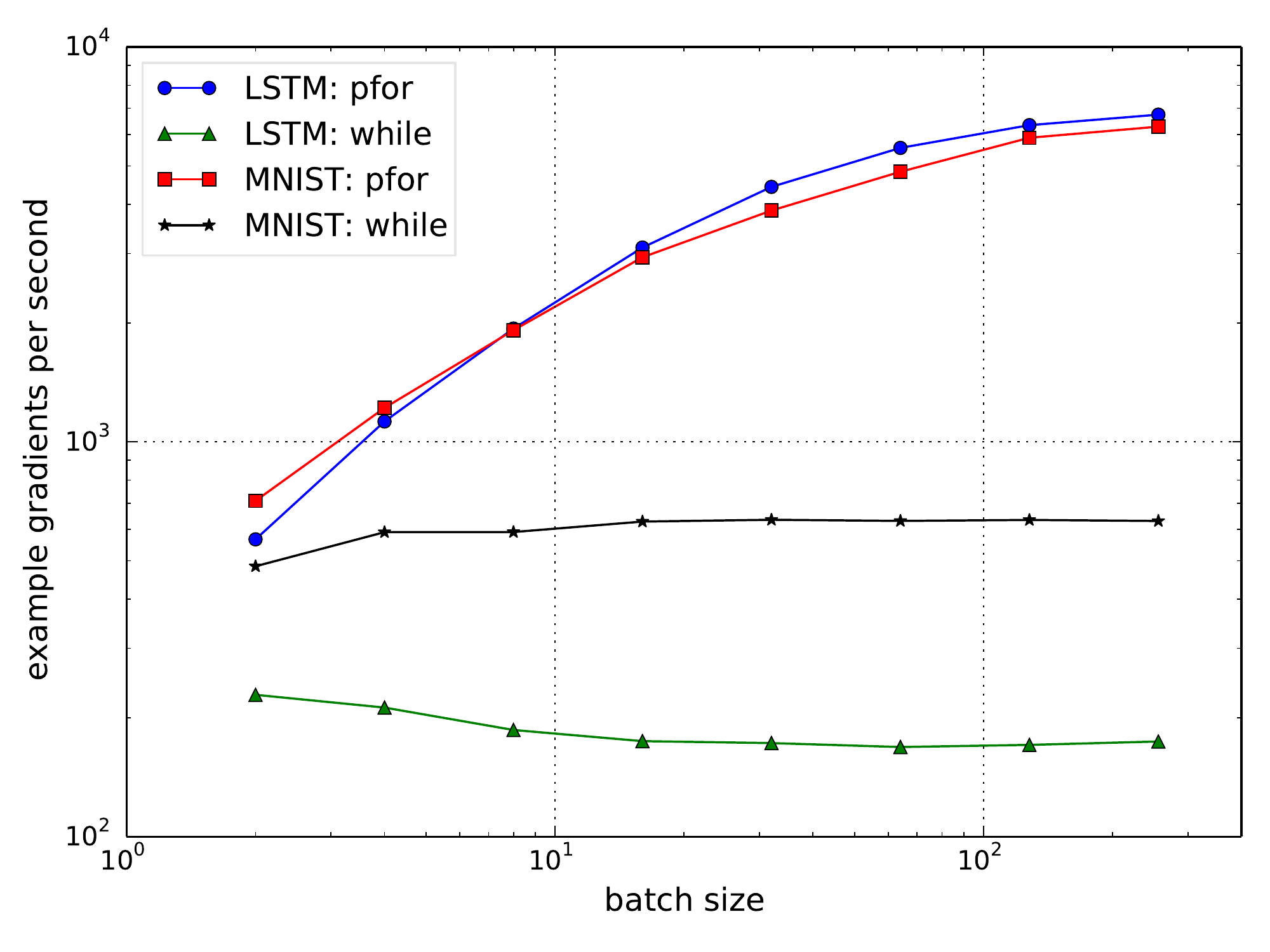}}
\vspace{-10pt}
\caption{Throughput of per-example gradient computation on MNIST and LSTM with and without auto-vectorization, on GPU. The x-axis is the number of examples in the batch. The y-axis is the number of per-example gradients that can be computed.}
\label{fig:per_eg}
\end{center}
\vskip -0.2in
\end{figure}

In this benchmark we evaluate the number of example gradients that \texttt{pfor} and \texttt{tf.while\_loop} based implementations can achieve versus the number of examples in the batch. Example is an image for MNIST and a complete input sequence for LSTM. All input sequences have length 10 in this benchmark. The \texttt{tf.while\_loop} based implementation simply iterates over all the examples in the batch, computing gradients for each. \texttt{pfor} based implementation vectorizes this iteration.

Figure \ref{fig:per_eg} shows that \texttt{tf.while\_loop} based implementation achieves almost constant throughput for both MNIST and LSTM models. On the other hand, \texttt{pfor} based implementation is able to utilize the GPU better by vectorizing the gradient computation. At the highest batch size of 256 for the LSTM model, \texttt{pfor} outperforms \texttt{tf.while\_loop} by a factor of 38.

\subsection{Jacobians}
\label{benchmark:jacobians}

\begin{figure}[t]
%\vskip 0.2in
\begin{center}
\centerline{\includegraphics[width=1.0\linewidth]
    {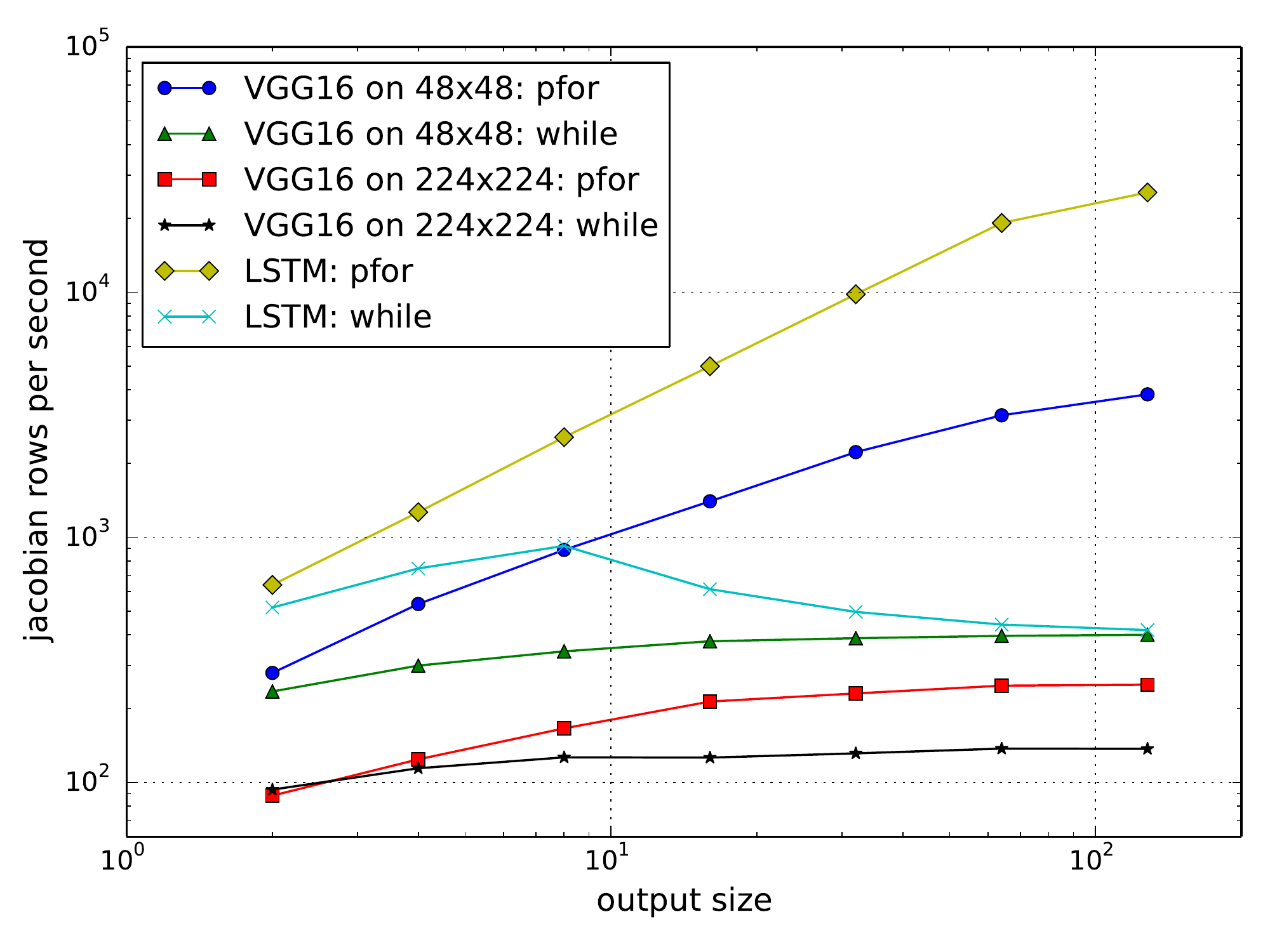}}
\vspace{-10pt}
\caption{Throughput of computing jacobians (output with respect to input) for VGG16 and LSTM models as output size is varied, with and without vectorization, on GPU. VGG16 uses two different input image sizes.}
\label{fig:jacobian}
\end{center}
\vskip -0.2in
\end{figure}

This benchmark looks at computing jacobians of model output with respect to model inputs on GPU.
As mentioned in \ref{jacobians}, TensorFlow currently has no efficient native support for computing jacobians. The best available option is to compute the gradients of each scalar in the output with respect to inputs. Each such gradient is a row of the jacobian matrix. We vary the output size and measure throughput
as rows of jacobian processed per second. This metric normalizes the compute done for a given task as we
vary the output size. Figure \ref{fig:jacobian} reports jacobian rows per second against output size.

Besides the LSTM model that we used in other benchmarks, we picked the VGG16 \cite{vgg} model 
since it supports different input image sizes as well as different number of output classes.
We report results for  two input image sizes: $48x48$ and $224x224$. The LSTM model is statically unrolled to 10 steps in this benchmark because of the TensorFlow's restriction described in \S\ref{jacobians}.

Figure \ref{fig:jacobian} shows that the unvectorized  implementation scales poorly with output sizes.
The only speed up it is able to get is from parallel execution of the loop body. \texttt{pfor} based implementation outperforms  \texttt{tf.while\_loop} based implementation by more than an order of magnitude at higher output sizes. At the largest output size of 128, \texttt{pfor} outperforms \texttt{tf.while\_loop} by over 60x.

\section{Summary}
We proposed applying static vectorization to dataflow IR like TensorFlow graphs. We implemented a
library that provides vectorized parallel-for loop construct for
TensorFlow. This enables applications ranging
from static auto-batching and per-example gradients, to jacobians, hessians and optimization of input pipelines.
GPU benchmarks show speedups of up to two orders of magnitude compared to TensorFlow's sequential loop and 
an order of magnitude against DyNet with dynamic batching. There is ongoing work on better handling loop variant shapes by automatic padding and masking.

Moving forward, these techniques can help optimize sequential loops. More research is needed in dealing
with sparse computation, recursive computation like tree or graph traversals,
handling memory constraints and applying polyhedral loop optimization,
generating more optimized code using better heuristics and hardware cost models, etc.

\bibliography{tfpaper}
\bibliographystyle{sysml2019}

\end{document}